\documentclass{article}
\usepackage[utf8]{inputenc}

\usepackage{amssymb}
\usepackage{amsmath}
\usepackage{graphicx}
\usepackage{authblk}

\usepackage{caption} 
\usepackage[subrefformat=parens,labelformat=parens]{subcaption} 

\def\uy {{\underline{y}}}
\def\uz {{\underline{z}}}
\def\uf {{\underline{f}}}

\title{Instability of oscillations \\
in the Rosenzweig-MacArthur model \\
of one consumer and two resources}

\author[1]{Przemysław Gawroński}
\author[2]{Alfio Borzì}
\author[1]{Krzysztof Kułakowski}

\affil[1]{Faculty of Physics and Applied Computer Science\\AGH University of Science and Technology\\al. Mickiewicza 30, Kraków, PL-30059, Poland}

\affil[2]{Institut für Mathematik, Universität Würzburg\\
Emil-Fischer-Strasse 30, Würzburg 97074, Germany}

\date{\today}

\begin{document}

\maketitle
\begin{abstract}
The system of two resources $R_1$, $R_2$ and one consumer $C$ is investigated within the Rosenzweig-MacArthur model with Holling type II functional response. The rates $\beta_i$ of consumption of resources $i=1,2$ are coupled by the condition $\beta_1+\beta_2=1$. The dynamic switching is introduced by a maximization of $C$: $d\beta_1/dt=(1/\tau) dC/d\beta_1$, where the characteristic time $\tau$ is large but finite. The space of parameters where both resources coexist is explored numerically. The results indicate that oscillations of $C$ and mutually synchronized $R_i$ which appear at $\beta_i=0.5$ are destabilized for $\beta_i$ larger or smaller. Then, the system is driven to one of fixed points where either $\beta_1>0.5$ and $R_1<R_2$ or the opposite. This behaviour is explained as an inability of the consumer to change the preferred resource, once it is chosen.

\end{abstract}

\section{Introduction}
The seminal papers by Lotka and Volterra \cite{LV1,LV2} initiated the research of modeling ecological processes \cite{mka0,ip}. 
Apart from the original system of one prey and one predator, the case of two preys and one predator plays a particular role as it is relatively simple and therefore can be investigated analytically \cite{c2r1,c2r2,c2r3,c2r4,c2r5}. \\

Here we are interested in the way how the predator (termed as consumer $C$ hereafter) distributes the rates of consumption of the preys (termed as resources $R_i$, $i=1,2$ hereafter). If $C$ and $R_i$ change in time, this distribution can also be time-dependent. In particular, these rates can be simply proportional to the amounts of resources $R_i$. For more complex cases, when this proportionality is absent, the term 'switching' has been coined \cite{mka}. \\

In a series of papers \cite{c2r1,c2r1a,c2r1b} various mathematical expressions 
(logistic, exponential) have been applied to model an optimal switching of the distribution of the rates of consumption between two resources. There, the rate of consumption 
$dC/dt$ has been maximized. This rate has been taken directly from the equation of motion. The optimization process has been applied immediately, i.e. each point of the trajectory has been optimized. In other words, the timescale of the optimization was assumed to be much shorter, than the timescale of the dynamics itself. On the other hand, the effect of delay of action of resources with respect to preys has been discussed in \cite{bnd1,bnd2}. There, the delay time was a fixed parameter.\\

Here we propose to use a direct and continuous drift of the rates as to increase $C$ (and not $dC/dt$) during the time evolution. As we have no direct expression for $C(t)$, our task is numerical. Further, we do not assume an instant optimization. Instead, the process of maximization is running with a finite speed, which is proportional to a perceived gain of the consumer. Namely, the larger is the increase of $C$ with the rate, the larger the speed is. Finally, the rates of the choice of this or that resource are coupled by the condition that the sum of these rates is constant. This condition can be understood as the limitation of time devoted to handling resources. Within this time, the consumer has a choice.

\section{Model equations}
Our starting point is the Rosenzweig-MacArthur set of equations \cite{zhf} for two resources $R_1$, $R_2$ and one consumer $C$
\begin{equation}
\frac{dC}{dt}= pC\frac{\beta_1R_1+\beta_2R_2}{1+b(\beta_1R_1+\beta_2R_2)} -mC 
\end{equation}
\begin{equation}
\frac{dR_1}{dt}=R_1(1-\alpha_{11}R_1-\alpha_{12}R_2)-\frac{pR_1\beta_1C}{1+b(\beta_1R_1+\beta_2R_2)}
\end{equation}
\begin{equation}
\frac{dR_2}{dt}=R_2(1-\alpha_{21}R_1-\alpha_{22}R_2)-\frac{pR_2\beta_2C}{1+b(\beta_1R_1+\beta_2R_2)}
\end{equation}
Here $\beta_i$ is the rate of taking advantage of resource $R_i$, and $\alpha_{ij}$ is the limitation of growth of resource $R_i$, imposed by resource $R_j$. Further, $p$ is the attack rate of the consumer, $b$ is the functional response term of the consumer, and $m$ is the mortality rate of the consumer \cite{zhf}.\\

The setting (1-3) is supplemented here by two more equations. First is the condition, that the rate of consumption is limited; the consumer has to make a choice between $R_1$ and $R_2$, hence
$\beta_2=1-\beta_1$. The second equation 
 \begin{equation}
\frac{d\beta_1}{dt}=\frac{1}{\tau}\frac{dC}{d\beta_1}
\end{equation}
is related to the decision of the consumer $C$. Namely, he intends to modify the coefficients $\beta_i$ as to get the maximal value of $C$. The time evolution of the $\beta$'s is assumed to be slower than this of $C$ and $R_i$, hence the time scale $\tau$ is longer than one. We note that with this method of optimization, the consumer can be stuck at a local maximum, which is not necessarily optimal. \\

\section{Calculations}

We would like to compute $\frac{dC}{d \beta_1}$. For this purpose, we make the following steps: 

1) We write our set of equations for $C$ and $R_1$, $R_2$ 
putting $\beta_2 = 1- \beta_1$. We also change notation in this section,
and write $\beta$ instead of $\beta_1$. We have 
\begin{equation}
\frac{dC}{dt}= pC \frac{\beta R_1+ (1- \beta) \, R_2}{1+b(\beta \, R_1+(1- \beta) \, R_2)} -mC 
\label{eRMAC}
\end{equation}
\begin{equation}
\frac{dR_1}{dt}=R_1(1-\alpha_{11}R_1-\alpha_{12}R_2)-\frac{pR_1\beta \, C}{1+b(\beta \, R_1+  (1- \beta) \, R_2)}
\label{eRMAR1}
\end{equation}
\begin{equation}
\frac{dR_2}{dt}=R_2(1-\alpha_{21}R_1-\alpha_{22}R_2)-\frac{pR_2  (1- \beta) \, C}{1+b(\beta \, R_1+  (1- \beta) \, R_2)} .
\label{eRMAR2}
\end{equation}

For our discussion, we can introduce the vector function 
$\uy=(C,R_1,R_2)$ and identify the system above with 
\begin{equation}
\uy'(t)= \uf (\uy, \beta) , \qquad \uy(t_0)=\uy_0 ; 
\label{eSysCompact}
\end{equation}
we denote the solution to this Cauchy problem with $\uy (t,t_0,\uy_0,\beta)$; 
we omit the other parameters since they are fixed. 

2) We recall a result concerning the dependence of the solution 
to \eqref{eSysCompact} on $\beta$. It is show that 
the vector function $\frac{d}{d \beta} \uy (t,t_0,\uy_0,\beta)$ solves the following Cauchy problem \cite{CoddingtonLevinson1955}
\begin{equation}
\frac{d}{d t} \, \uz  = \partial_y \uf (\uy,\beta) \, \uz 
+ \frac{d}{d \beta}  \, \uf (\uy ,\beta) ,
\label{eSysZ}
\end{equation}
with the initial value $\uz (t_0)=0$. We denote with $\partial_y \uf (\uy,\beta) $ 
the Jacobian of $\uf$ (the r.h.s. of \eqref{eRMAC} - \eqref{eRMAR2}) 
with respect to $\uy=(C,R_1,R_2)$.

Therefore we obtain $\frac{dC}{d \beta} (t) = z_1(t)$. Consequently, we
obtain a differential equation for $\beta$: 
 \begin{equation}
\frac{d\beta}{dt}=\frac{1}{\tau}\frac{dC}{d\beta} = \frac{z_1}{\tau}. 
\label{eBeta}
\end{equation}

Notice we have to compute seven variables $(C,R_1,R_2,z_1,z_2,z_3,\beta)$ 
by solving the coupled system 
\eqref{eRMAC}, \eqref{eRMAR1}, \eqref{eRMAR2}, 
\eqref{eSysZ} and \eqref{eBeta}, with given initial 
conditions $(C(t_0),R_1(t_0),R_2(t_0),\beta(t_0))$, and 
$z_j(t_0)=0$, $j=1,2,3$. \\


Two hints can be added to limit the space of parameters. First, in the lack of consumers (here: if $C=0$) there is a non-zero fixed point: 
\begin{eqnarray}
R_1^*=\frac{\alpha_{22}-\alpha_{12}}{\alpha_{11}\alpha_{22}-\alpha_{12}\alpha_{21}}\\
R_2^*=\frac{\alpha_{11}-\alpha_{21}}{\alpha_{11}\alpha_{22}-\alpha_{12}\alpha_{21}}
\end{eqnarray}
To keep $R_i^*$ positive, we should assume either diagonal elements $\alpha_{ij}$ larger than non-diagonal ones, or the opposite; our choice is the former. Second, for one pair 'consumer-resource', the stability of a limit cycle demands that $p-mb>2(b+1)/p$. As we are interested in limit cycles, we keep $p=3.0$, $b=2.0$ and $m<0.5$, if not stated otherwise.\\

Equations (1-3) are solved with conventional Runge-Kutta 4 method. Equation (5) needs some care, because we have no analytical function 
$C(\beta)$. Therefore at each time step we calculate $C(\beta+\epsilon)$ and $C(\beta-\epsilon)$. Then the new value of $\beta$ is
\begin{equation}
\beta(t+dt)=\beta(t)+\frac{C(\beta+\epsilon)-C(\beta-\epsilon)}{2\epsilon}\frac{dt}{\tau}
\end{equation}
Typically, $\epsilon=5\times10^{-6}$, if not stated otherwise. Also $m=0.2$, $p=3.0$, $b=2.0$
$\alpha_{12}=\alpha_{21}=0.65$, and $\alpha_{11}=\alpha_{22}$. In the next section we concentrate on the role of $\alpha_{ii}$, $\tau$ and $\beta_1(t=0)$.

\section{Results}

\begin{figure}[hb]
\includegraphics[width=1.0\textwidth]{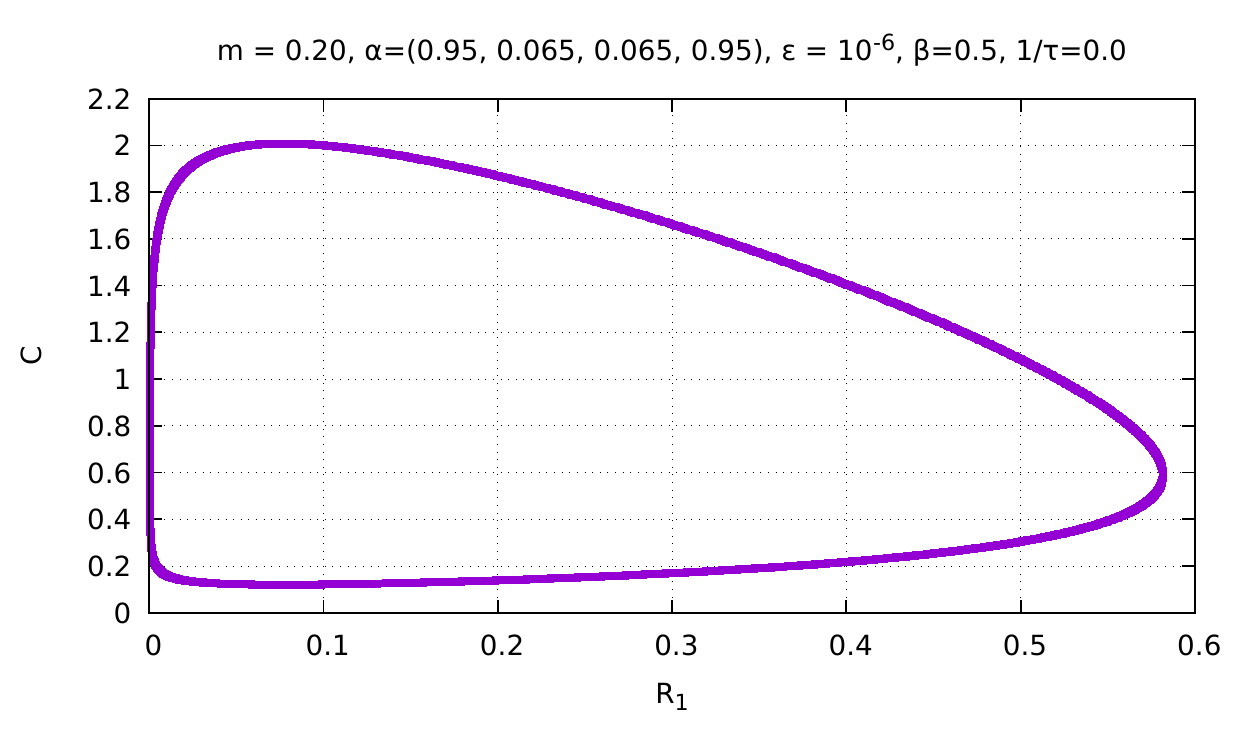}
\caption{Phase A: the loop $C vs. R_1$}
\label{fig1}
\end{figure}

If, as noted above, the matrix $\alpha_{ij}$ is symmetric, the transformation $\beta_i\to \beta_{3-i}$ should not change the outcome; only the values of $R_i$ are interchanged. As a consequence, at the point $\beta_i=0.5$ the derivative $\delta C/\delta \beta_i=0$, hence $d\beta_i/dt=0$. There we observe oscillations of $C$ and of $R_1=R_2$; both resources perfectly synchronize and behave as one. The loop C vs R is shown in Fig. \ref{fig1}. This phase will be termed as phase A from now on.\\

\begin{figure}[ht]
\includegraphics[width=1.0\textwidth]{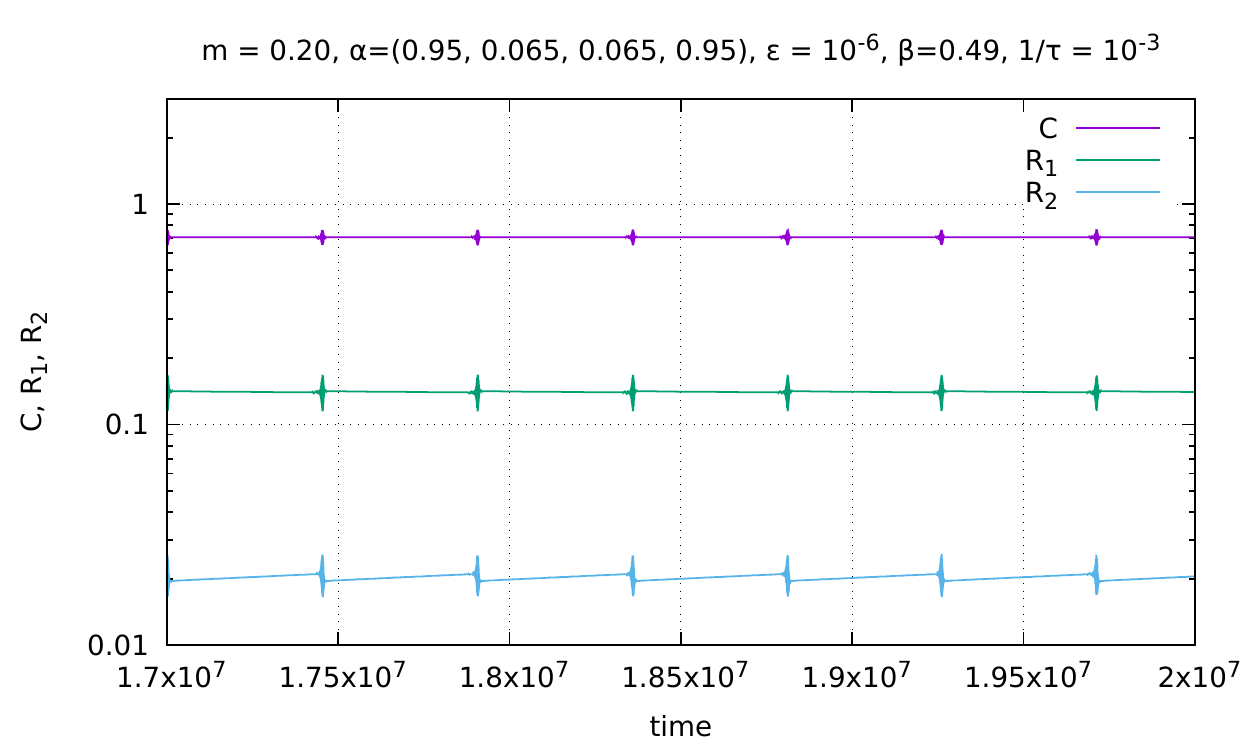}
\caption{Time dependence of $C, R_1, R_2$ in phase B}\label{fig2log}
\end{figure}

A small shift, about 0.01, of the initial value of $\beta_1$ from the symmetric value 0.5 drives the system to a different state, which itself depends on $\tau$. For values of $1/\tau$ small enough, the absolute value of the difference $|\beta_i-0,5|$ increases to about 0.03 and remains there, with small periodic oscillations. The stationary behaviour of $C$ and $R_i$, termed here as phase B, is shown in Fig. \ref{fig2log}.\\

\begin{figure}[ht]
\includegraphics[width=1.0\textwidth]{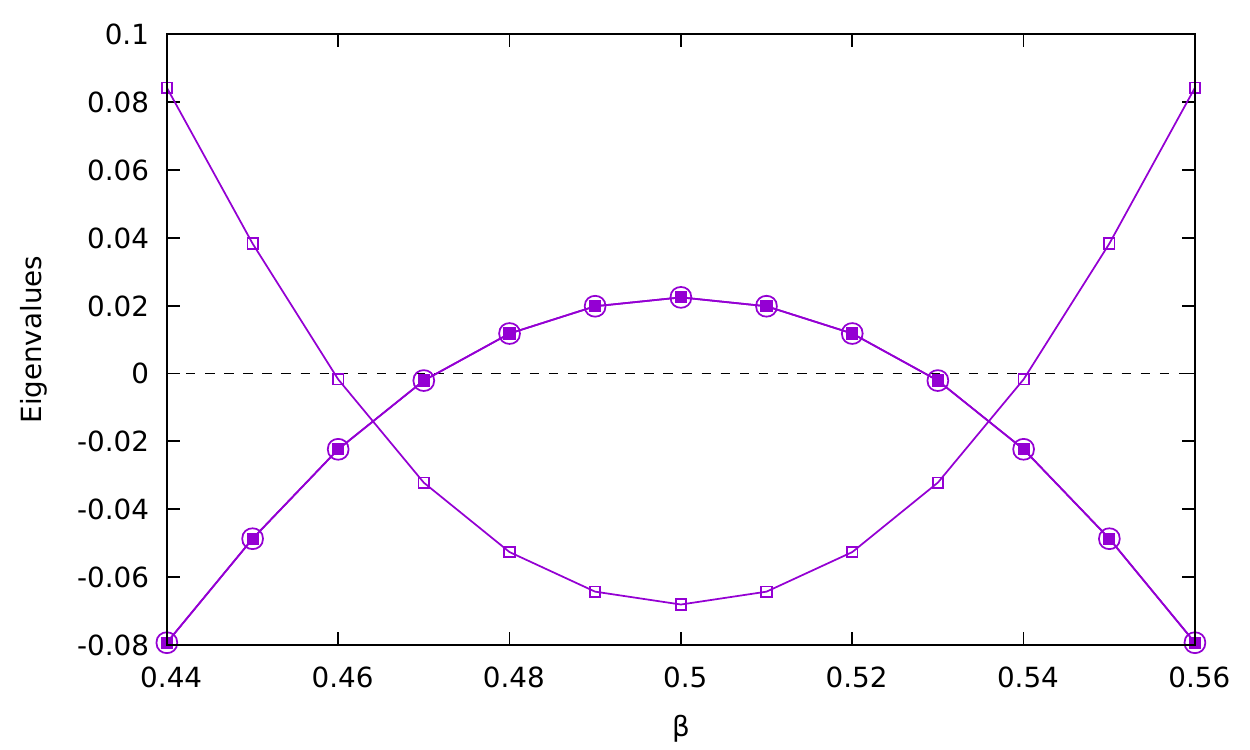}
\caption{The real parts of three eigenvalues of the Jacobian for $\tau=\infty$. The concave curve is doubly degenerated. Two stable areas are visible near $\beta_i=0.465$ and $0.535$}\label{fig5}
\end{figure}

The necessary condition of the appearance of phase B is the stability of this phase for $1/\tau=0$. In Fig. \ref{fig5} we show the real parts of the eigenvalues of the related Jacobian in three-dimensional space $C,R_1,R_2$. They appear to be negative in two narrow windows near $\beta_i=0.465$ and $0.535$. Details on these fixed points are given in the Appendix.\\

It is characteristic that for $\beta_1>0.5$, the resource $R_1$ is less than $R_2$, and the opposite. This result indicates, that the consumer maintains her/his initial preference, imposed by the choice of the initial condition on $\beta_i$. This is so despite the fact, that the more exploited resource is less abundant. On the other hand, the variable $C$ in phase B (about 0.71) is slightly larger than the time average of $C$ in phase A (about 0.68). \\ 

\begin{figure}
\includegraphics[width=1.0\textwidth]{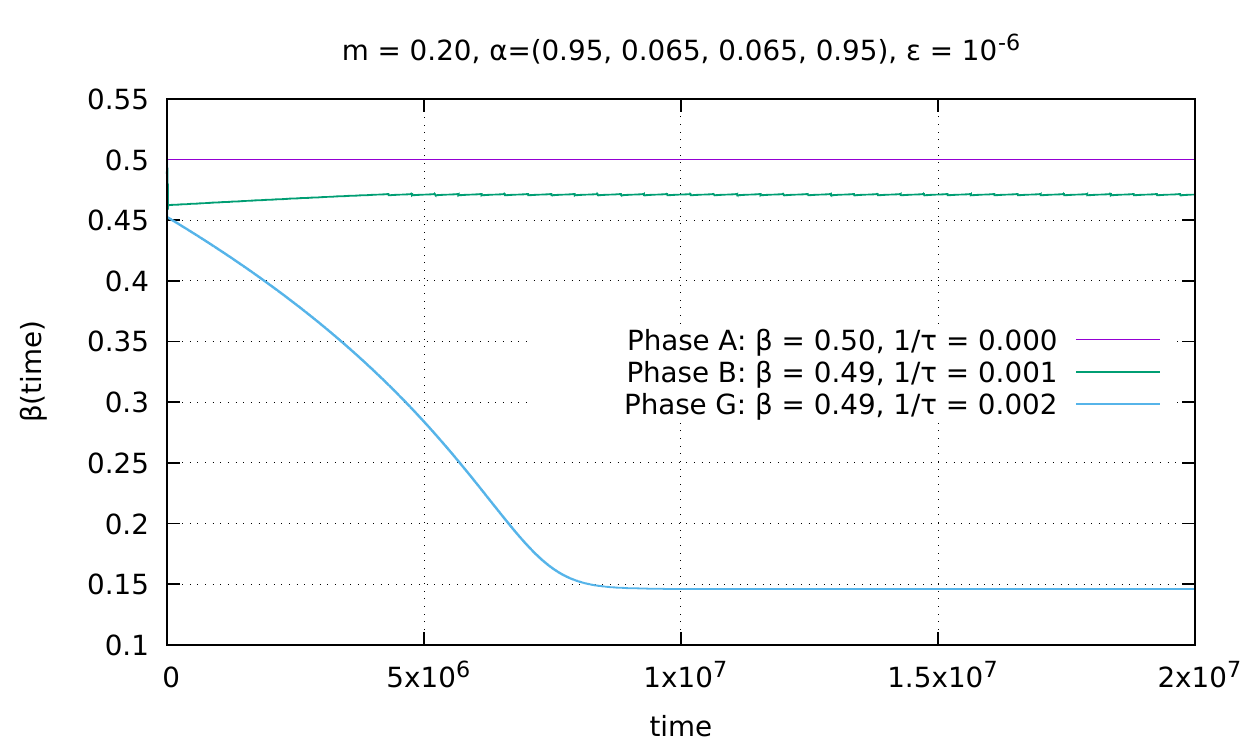}
\caption{$\beta(t)$ in three phases: A, B and G, from top to bottom.}\label{fig3}
\end{figure}

\begin{figure}
\subcaptionbox{$\beta(t)$ in phase B. 
\label{fig3a}}{\includegraphics[width=0.7\textwidth]{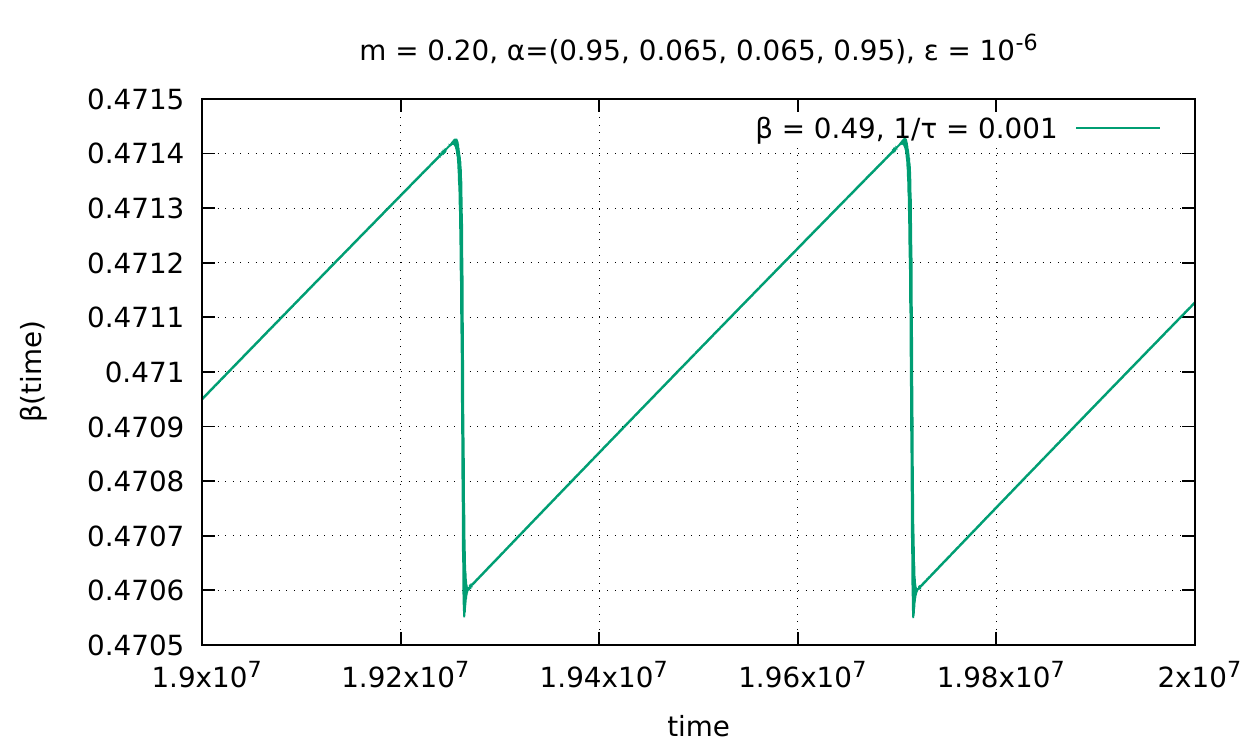}}
\subcaptionbox{ Larger zoom on $\beta(t)$ in phase B. \label{fig3x}}{\includegraphics[width=0.7\textwidth]{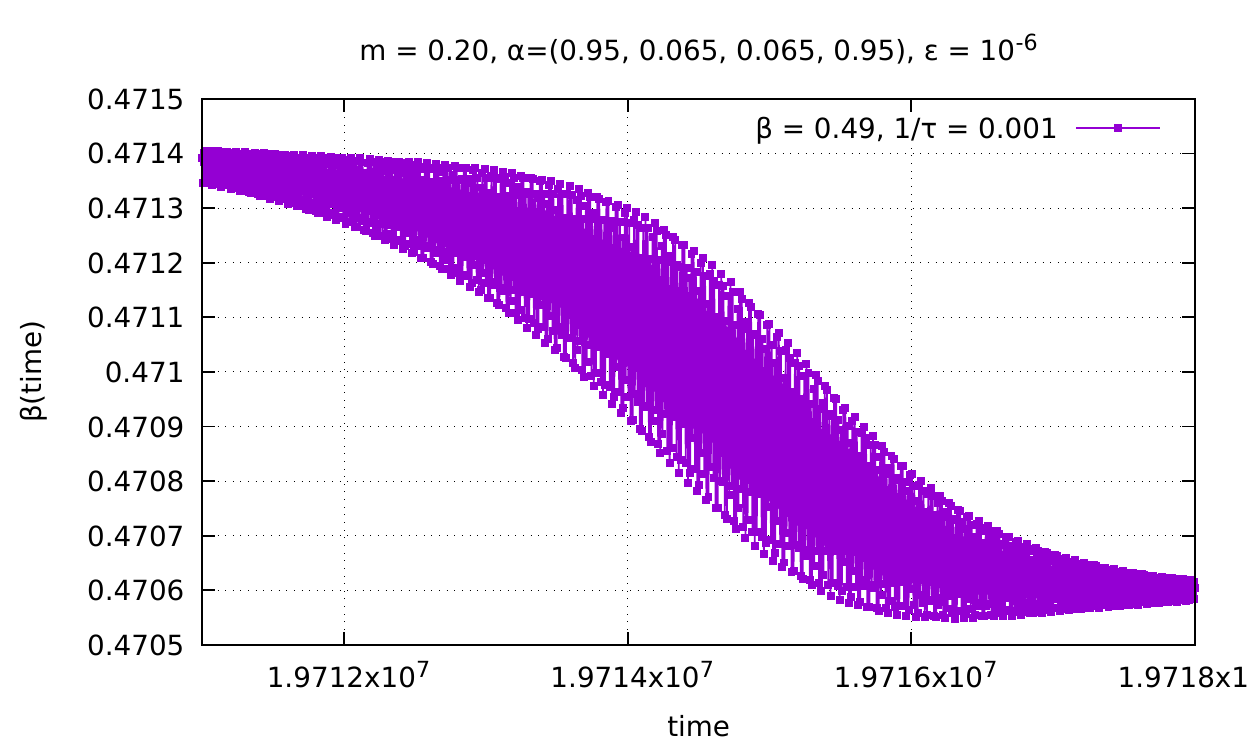}}
\subcaptionbox{Even larger zoom on the falling part of $\beta(t)$ in phase B. \label{fig3b}}{\includegraphics[width=0.7\textwidth]{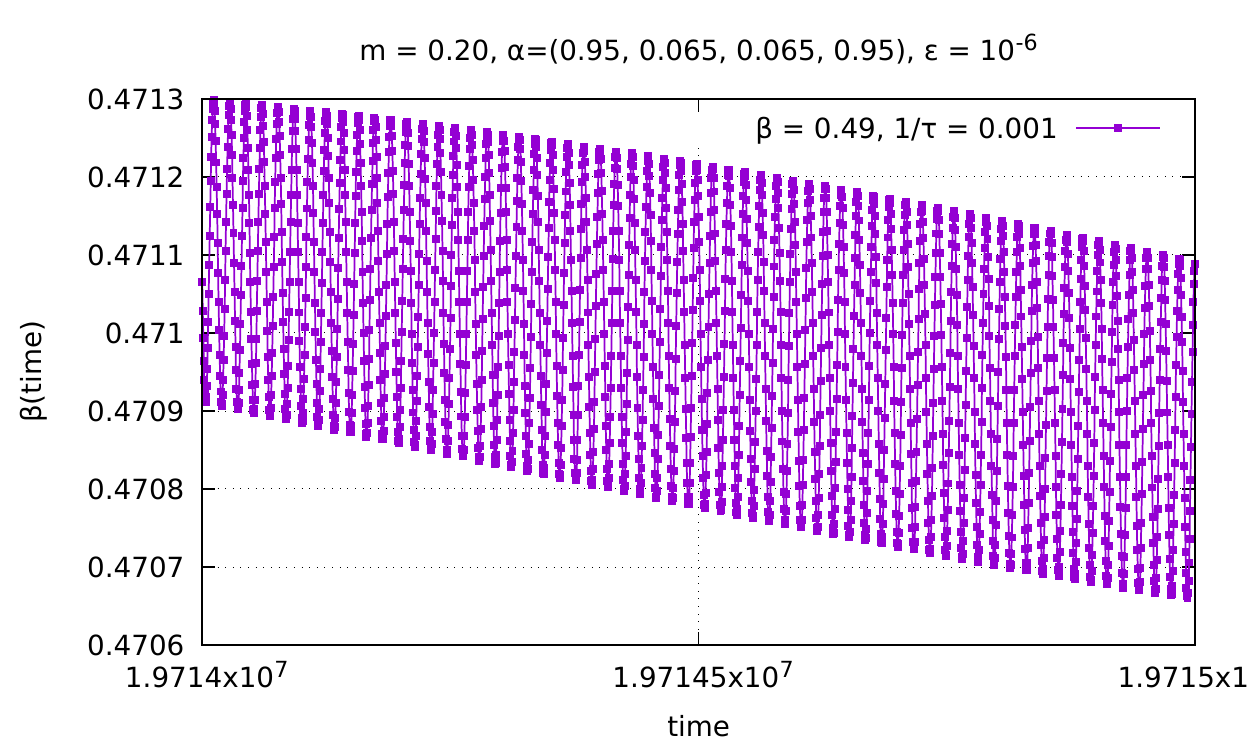}}
\caption{Time dependence of $\beta_1$. }
\end{figure}

\begin{figure}
\subcaptionbox{$C\; vs. \beta$ in phase B. 
\label{fig3d}}{\includegraphics[width=0.7\textwidth]{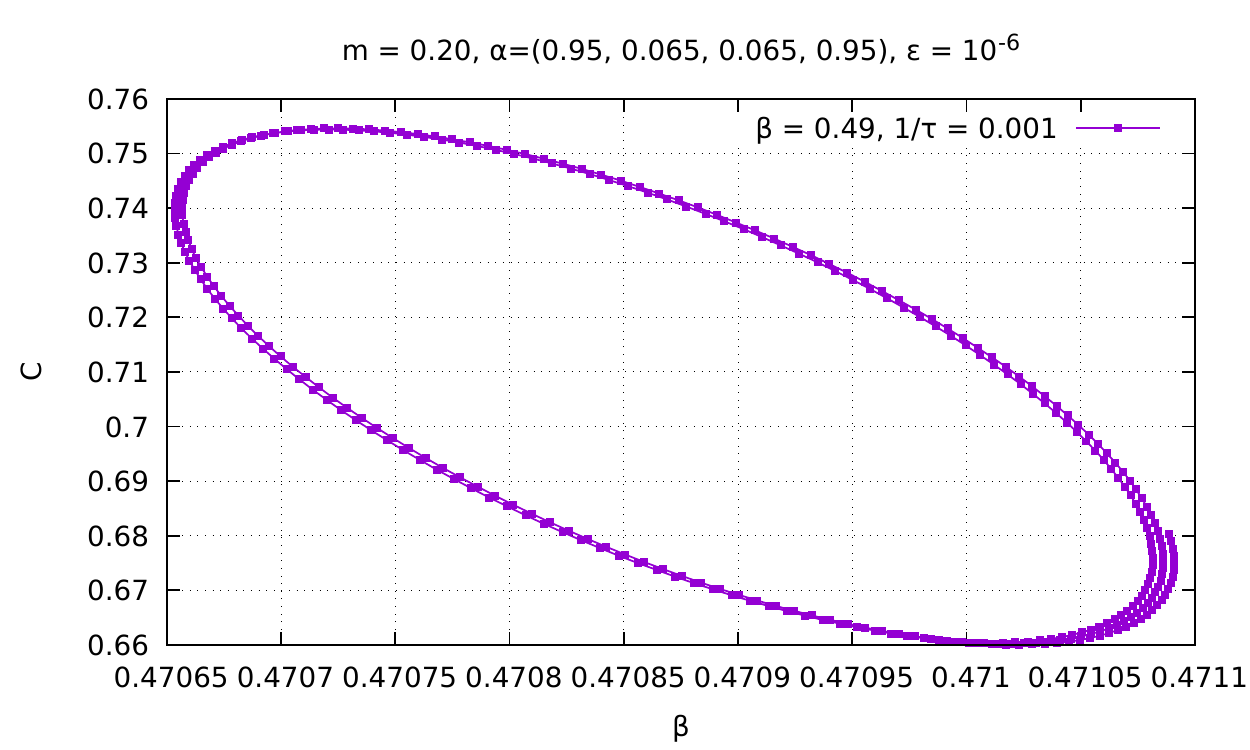}}
\subcaptionbox{ $R_1 \;vs. \beta$ in phase B. \label{fig3e}}{\includegraphics[width=0.7\textwidth]{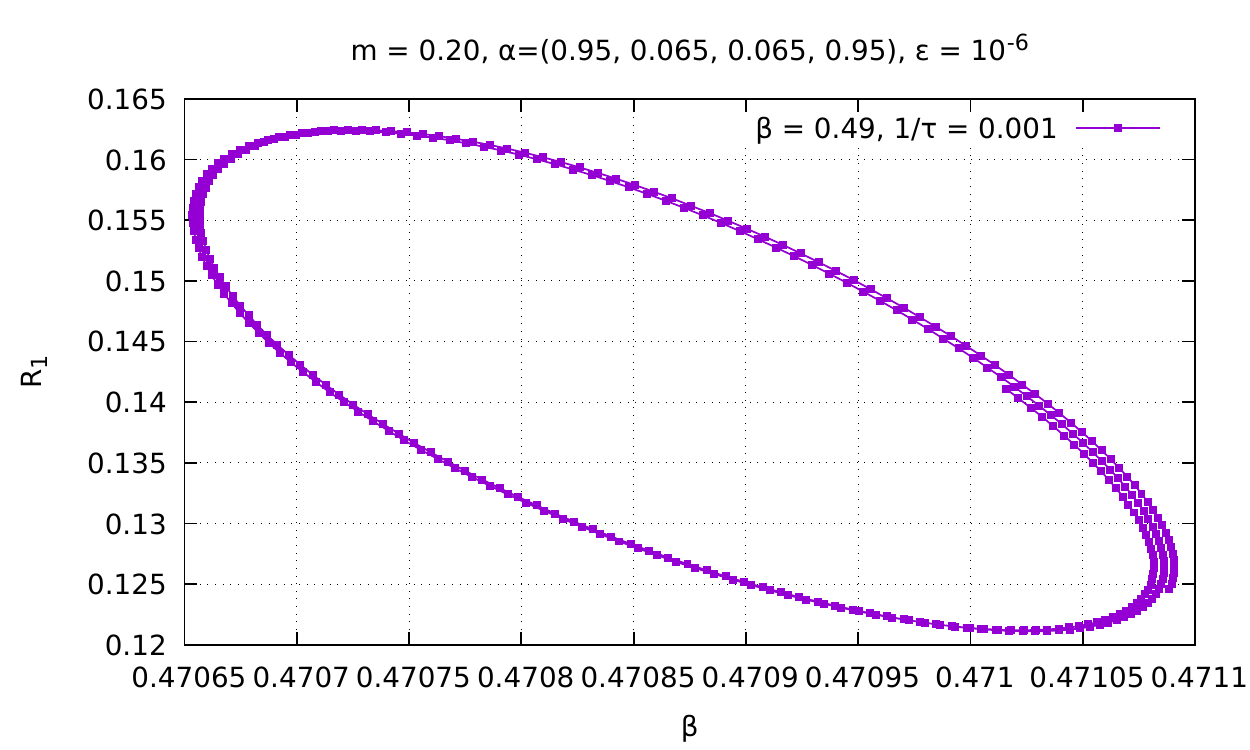}}
\subcaptionbox{ $R_2 \; vs. \beta$ in phase B. \label{fig3f}}{\includegraphics[width=0.7\textwidth]{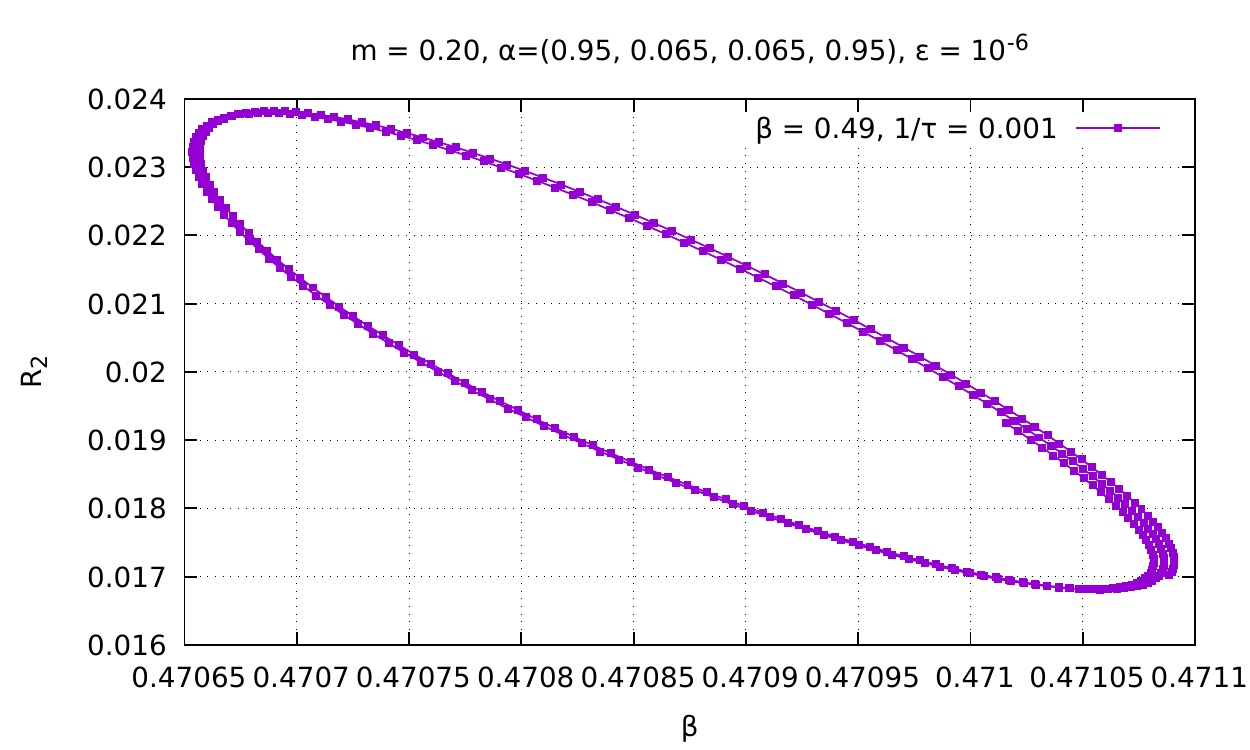}}
\caption{The dependencies of $C$, $R_1$, $R_2$ on $\beta_1$ in phase B during the oscillating phase. The time window is taken from the center of Fig.\ref{fig3b}. }
\end{figure}

The small oscillations of $\beta_i$ in phase B, barely visible in Fig. \ref{fig3} and expanded in Figs. \ref{fig3a}, \ref{fig3x} and \ref{fig3b}, are a consequence of finite $\tau$, as their period is about to be doubled when $\tau$ increases five times. The behaviour is remarkable:  $\beta_1$ increases linearly, then falls into oscillations, dropping down, then again increases and so on. This behaviour of $\beta_i$ induces similar variations of $C$ and $R_i$ (Fig. \ref{fig2log}). In Figs. \ref{fig3d}, \ref{fig3e} and \ref{fig3f} the dependences $C(\beta)$, $R_1(\beta)$ and $R_2(\beta)$ are shown, within the time range where the amplitudes of the oscillations are relatively large (as near the center of Fig. 5b). There we see the loops at each dimension.\\

For the parameter $1/\tau$ above some critical value $1/\tau*$, the more exploited resource $R_i$ entirely vanishes. Still, the consumer $C$ maintains her/his initial preference; once $R_2=0$, $\beta_1$ decreases to about 0.15, and if $R_1=0$, $\beta_1$ increases to about 0.85. On the other hand, in both cases $C$ increases to about 1.4. This phase is termed G from now on. \\
 
\begin{figure}
\includegraphics[width=1.0\textwidth]{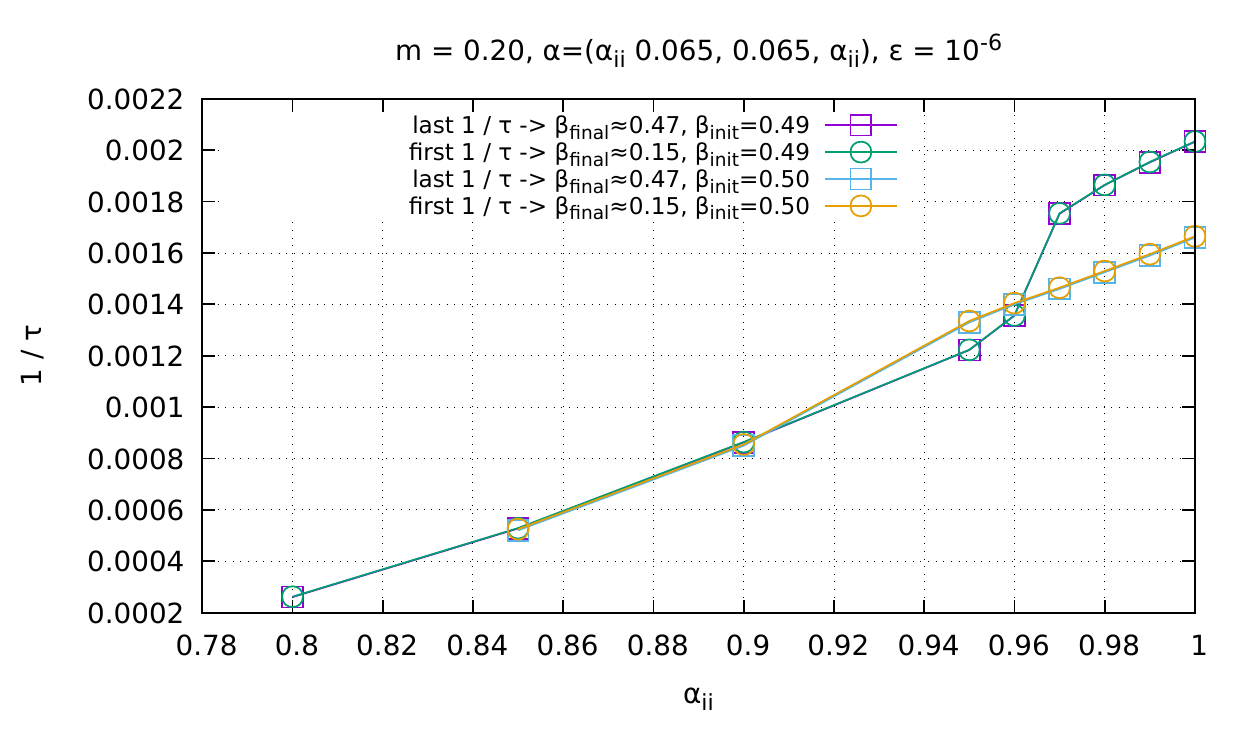}
\caption{The boundary between phases B and G, with G above the plot.  }\label{fig4}
\end{figure}

\begin{figure}
\subcaptionbox{$R_2(t)$ near the transition - the system remains in phase B \label{fig8a}}{\includegraphics[width=1.0\textwidth]{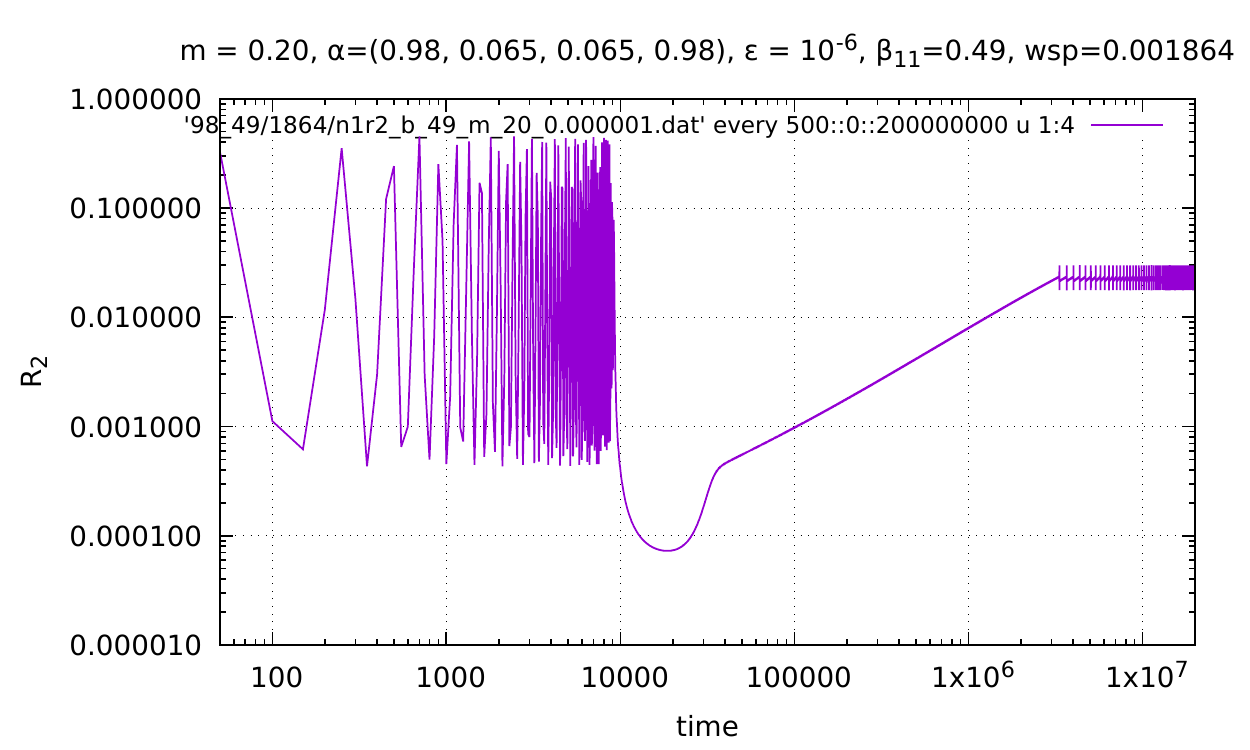}}
\subcaptionbox{ $R_2(t)$ near the transition - the system switches to phase G \label{fig8b}}{\includegraphics[width=1.0\textwidth]{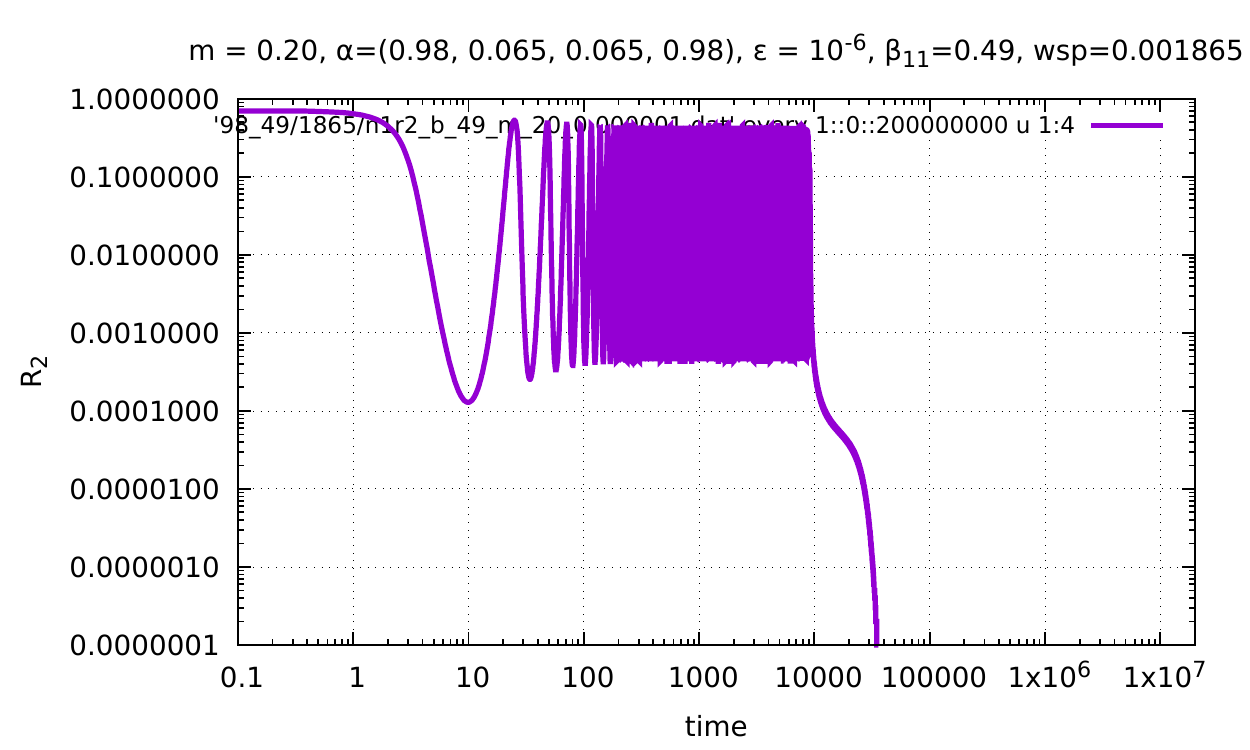}}
\caption{Two trajectories $R_2(t)$ at the boundary between phases B and G. }
\end{figure}

 The critical values of $1/\tau*$ at the boundary between the phases B and G are shown in Fig. \ref{fig4}, for different values of $\alpha_{ii}$. Actually, the transition to phase G is caused by a collision of the trajectory $R_i(t)$ with the invariant manifold $R_i(t)=0$. As shown in Figs. \ref{fig8a} and \ref{fig8b}, this collision can happen at a transient stage, and it is not easy to disentangle the deterministic character of this transition. Despite of this difficulty, the boundary between the phases is quite smooth. \\

 Further, a difference of $\tau*$ appears between the plots for $\beta_i=0.5$ and for its other values. This difference can be attributed to stronger transient oscillations of $R_i$'s, observed for the state where $\beta_i=0.5$. \\

If we start simulations from the state where $R_2=0$, obviously the symmetry of an interchange $\beta_1-\beta_2$ is lost. In this case, the system is driven to the phase G for any initial value of $\beta_1$, if only $\tau$ is finite. The case of infinite $\tau$ is an exception; there, the time evolution of $\beta_{i}$ is blocked, and the problem is reduced to the standard case of one consumer and one resource \cite{ip}.


\section{Discussion}

The phase A can be considered to be close to the limit $\tau \to \infty$, where the evolution of coefficients $\beta_i$ can be neglected. The synchronization of $R_1$ and $R_2$, observed in this phase, effectively reduces the problem to the case of one consumer and one resource. As it was demonstrated in \cite{ip}, the solution is a stable limit cycle. This means in particular, that the fixed point at $\beta_i=0.5$ is unstable. The same cycle is observed in phase A, if the initial values of the rates $\beta_i$ are equal to 1/2. As the whole problem is symmetric with respect to an interchange of $\beta_1$ and $\beta_2$, and $\beta_1=1-\beta_2$, the derivative $dC/d\beta_i=0$ at this point, and the time evolution of the coefficients $\beta_i$ is blocked. In the phase B this symmetry is broken, as the coefficients $\beta_i$ are different.\\

  Some analogy can be drawn with the classical Landau theory of phase transitions \cite{nlt8}. There, suppose that $\Delta=\beta-1/2$, and $C(\beta)=e\Delta^2-f\Delta^4$ (with $e,f$ - positive constants) plays the role of a negative potential which is maximized in a stationary state. Within this setting, $\Delta=0$ is a meta-stable state A, at a local minimum of $C(\Delta)$. The phase B appears to be stable, with the non-zero value of $\Delta$. However, this analogy remains incomplete, because actually the problem is 4-dimensional, with variables $C$, $R_1$, $R_2$ and $\beta_1$. The fixed point at $\beta_1=0.5$ in the space ($C,R_1,R_2$) is unstable and a limit cycle appears there, in accordance with literature. Further, once the dynamics of $\beta_1$ is enabled, this limit cycle is unstable itself. \\
  
 The dynamics of $\beta_i(t)$, shown in Figs. \ref{fig3a}, \ref{fig3x} and \ref{fig3b} indicates that the periodic trajectory contains a part wounded onto a torus and a part where the time dependence is linear. This picture can be seen as a periodic orbit with two successive Hopf bifurcations \cite{claus}, where the driving force along the orbit is due to finite speed $d\beta_i/dt$.\\
 
 The phase G - a stable fixed point - appears for smaller values of the characteristic time $\tau$. We observe that its stability is not a consequence of the time evolution of $\beta_i$, as this phase is stable also in the limit of $\tau$ infinite. The role of the evolution is to drive $\beta_1$ to a final value (from any one less than $0.5$ to about $0.15$), passing from the phase B to the phase G. We add that for $\tau$ infinite, $R_2=0$ and $\beta_1$ larger than some critical value (about $0.62$ for  $\alpha_{11}=0.95$, other parameters as above) the fixed point of the phase G is unstable and periodic oscillations of $C$ and $R_1$ are restored.\\
 
 Summarizing the results, in the presence of the evolution of the coefficients $\beta_i$ the relatively large oscillations of the variables $C$ and $R_i$ in the phase A appear to be unstable, and they persist only at $\beta_i=0.5$. The finite speed $1/\tau$ is supposed to drive the system to larger value of $C$. However this evolution keeps the system at the smaller $R_i$, leaving less exploited the resource which is more abundant. The difference between the resources $R_i$ is caused just by the difference of their exploitation, measured by the coefficients $\beta_i$. As the manifold $\beta_i=0.5$ is invariant, the consumer is not able to switch to the larger resource. For even shorter time $\tau$, the system is driven from the phase B to a stable fixed point in the phase G, where one of the resources $R_i$ vanishes.  \\


\section{Appendix: the fixed points for $\tau=\infty$}

For non-zero values of $C$, $R_1$, $R_2$, and for $\beta_2=1-\beta_1$ independent on time, the equations for the fixed point are as follows:

\begin{eqnarray}
\alpha_{11}R_1+\alpha_{12}R_2+\phi \beta_1 C&=&1\nonumber\\ 
\alpha_{21}R_1+\alpha_{22}R_2+\phi \beta_2 C&=&1\nonumber\\
\beta_1R_1+\beta_2R_2&=&w\nonumber
\end{eqnarray}
where $\phi=(p-mb)/r$, and $w=m/(p-mb)$. The related determinants are 
\begin{eqnarray}
D&=&\phi[(\alpha_{21}+\alpha_{12})\beta_1\beta_2-\alpha_{22}\beta_1^2-\alpha_{11}\beta_2^2]\nonumber\\
D_1&=&\phi(\beta_1\beta_2+w\alpha_{12}\beta_2-w\alpha_{22}\beta_1-\beta_2^2)\nonumber\\
D_2&=&\phi(\beta_1\beta_2+w\alpha_{21}\beta_1-w\alpha_{11}\beta_2-\beta_1^2)\nonumber\\
D_3&=&w(\alpha_{11}\alpha_{22}-\alpha_{12}\alpha_{21})+\beta_1(\alpha_{12}-\alpha_{22})+\beta_2(\alpha_{21}-\alpha_{11})\nonumber
\end{eqnarray}
and the solution is $R_1=D_1/D$, $R_2=D_2/D$, and $C=D_3/D$. Analysis of signs of the eigenvalues of the Jacobian indicates, that this fixed point is unstable in the whole range of $\beta_1$, except two narrow bands near $\beta_i=0.47$ and $0.53$. These bands are shown in Fig. \ref{fig5}. This indicates, that the phase B is stable also for $1/\tau=0$.\\

There are also two other fixed points, one with $R_1=0$ and another with $R_2=0$. For the latter, $R_1=m/(\beta_1(p-mb))$, and $C=r(1-\alpha_{11}R_1)(1+b\beta_1R_1)/(p\beta_1)$. Here the analysis of the eigenvalues of the Jacobian shows that a stable area for $R_2=0$ appears in the range approximately $0.08<\beta_1<0.46$, both for $\alpha_{11}=0.85$ and $0.95$. \\

The results for infinite $\tau$ indicate that the fixed points are not isolated, but form a kind of bands. When for finite $\tau$ the evolution of $\beta_i$ is enabled, small variations of the related variables are observed in phase B, whereas in phase G the solution is a stable fixed point.


\begin{thebibliography}{99}
\bibitem{LV1} A. J. Lotka, Elements of Physical Biology, Williams and Wilkins, 1925.

\bibitem{LV2}  V. Volterra, Variazioni e fluttuazioni del numero d'individui in specie animali conviventi, Mem. Acad. Lincei Roma. 2 (1926) 31–113.

\bibitem{mka0} From Energetics to Ecosystems: The Dynamics and Structure of Ecological Systems, edited by N. Rooney, K. S. McCann and D. L. G. Noakes, Springer, Dordrecht 2007.

\bibitem{ip} M. Iannelli and A. Pugliese, An Introduction to Mathematical Population Dynamics: Along the trail of Volterra and Lotka, Springer 2014.

\bibitem{c2r1} V. Krivan, Optimal foraging and predator-prey dynamics, Theoretical Population Biology, 49 (1996) 265-290.

\bibitem{c2r2} E. Sanchez-Palencia and J.-P. Francoise, Topological remarks
and new examples of persistence of diversity in biological
dynamics, Discrete and Continuous Dynamical Systems Series
12/6 (2019) 1775-1789.

\bibitem{c2r3} D. Savitri, A. Suryanto, W. M. Kusumawinahyu and Abadi, 
Dynamics of two preys --- one predator system with competition 
between preys, J. Phys. Conf. Series, 1562 (2020) 012010.

\bibitem{c2r4} K. Manna, V. Volpert and M. Banerjee,
Dynamics of a diffusive two-prey-one-predator model with nonlocal intra-specific competition for both the prey species,
Mathematics, 8 (2020) 101.

\bibitem{c2r5} A. Yamauchi, Y.Ikegawa, T. Ohgushi and  T. Namba, Density regulation of co-occurring herbivores via 
two indirect effects mediated by biomass and 
non-specific induced plant defenses, Theoretical Ecology 14, (2021) 41-55.

\bibitem{mka} M. Koen-Alonso, A process-oriented approach to the multispecies functional response, in \cite{mka0}, pp. 1-36.

\bibitem{c2r1a} V. Krivan and A. Sikder, Optimal foraging and predator-prey dynamics II, Theoretical Population Biology, 55 (1999) 111-126.

\bibitem{c2r1b} V. Krivan and J. Eisner, Optimal foraging and predator-prey dynamics III, Theoretical Population Biology, 63 (2003) 269-279.

\bibitem{bnd1} B. Barman and B. Ghosh,
Dynamics of a spatially coupled model with delayed prey dispersal,
International Journal of Modelling and Simulation, 2021
(doi:10.1080/02286203.2021.1926048)

\bibitem{bnd2} B. Barman and B. Ghosh,
Role of time delay and harvesting in some predator–prey communities with different functional responses and intra-species competition,
International Journal of Modelling and Simulation, 2021 (https://doi.org/10.1080/02286203.2021.1983747)



\bibitem{zhf} Z. Hajian-Forooshani and J. Vandermeer, Viewing communities as coupled oscillators: elementary forms from Lotka and Volterra to Kuramoto, Theoretical Ecology 14(12) (2021).

\bibitem{CoddingtonLevinson1955} 
E. A. Coddington and N. Levinson, 
Theory of Ordinary Differential Equations, McGraw-Hill, 1955. 

\bibitem{nlt8} W. Nolting, Theoretical Physics 8: Statistical Physics, Springer, Cham 2018, p. 308.

\bibitem{claus} J. Claus, M. Ptashnyk, A. Bohmann and A. Chavarria-Krauser, Global Hopf bifurcation in the ZIP regulatory system, J. Math. Biol. 71 (2015) 795.

\end{thebibliography}
\end{document}